\documentclass[pdflatex,sn-mathphys-num]{sn-jnl}

\usepackage{graphicx}%
\usepackage{multirow}%
\usepackage{amsmath,amssymb,amsfonts}%
\usepackage{amsthm}%
\usepackage{mathrsfs}%
\usepackage[title]{appendix}%
\usepackage{xcolor}%
\usepackage{textcomp}%
\usepackage{manyfoot}%
\usepackage{booktabs}%
\usepackage{algorithm}%
\usepackage{algorithmicx}%
\usepackage{algpseudocode}%
\usepackage{listings}%
\usepackage{booktabs} 
\usepackage{siunitx}
\usepackage{aas_macros}

\usepackage{color}
\definecolor{orange}{rgb}{1,0.5,0}

\begin{document}

\title[Particle dynamics in TOI-178 planetary system ]{Particle dynamics in TOI-178 planetary system }

\author[1,4]{\fnm{Jovan} \sur{Boskovic}}\email{jovan.boskovic@iag.uni-stuttgart.de}

\author*[2,1,3]{\fnm{Rafael} \sur{Sfair}}\email{rafael.sfair@unesp.br}

\author[1]{\fnm{Christoph M.} \sur{Schäfer}}\email{ch.schaefer@uni-tuebingen.de}

\affil[1]{\orgname{Institute for Astronomy and Astrophysics, Department of Computational Physics, University of Tübingen}, \orgaddress{\street{Auf der Morgenstelle 10}, \city{72076 Tübingen}, \country{Germany}}}

\affil[2]{\orgname{UNESP - São Paulo State University}, \orgaddress{\street{Av. Ariberto Pereira da Cunha, 333}, \city{Guaratinguetá}, \postcode{12516-410}, \state{São Paulo}, \country{Brazil}}}

\affil[3]{\orgname{LIRA, Observatoire de Paris, Université PSL, Sorbonne Université, Université Paris Cité, CY Cergy Paris Université, CNRS}, \orgaddress{\street{5 Place Jules Janssen}, \city{Meudon}, \postcode{92190}, \state{Île-de-France}, \country{France}}}

\affil[4]{\orgname{Institute of Aerodynamics and Gas Dynamics, University of Stuttgart}, \orgaddress{\street{Wankelstraße 3}, \city{70563 Stuttgart}, \country{Germany}}}


\abstract{
The TOI-178 system hosts six planets with five of them locked in a
2:4:6:9:12 Laplace resonance chain. We perform N-body simulations 
to investigate the dynamics of test particles in this system. We
{observe} that co-orbital regions around each planet are approximately 30\% wider
than predicted by classical theory for planets in the resonance chain, while
TOI-178b, which lies outside the chain, shows a 52\% enhancement. The region
between TOI-178e and TOI-178f reveals Kirkwood gap-like structures created by
mean-motion resonances with TOI-178f (4:3, 5:4, 6:5) and TOI-178g (5:3), where
particle clearing occurs on 500-year timescales. An extended integration of the
innermost region (0.015-0.025 au) shows periodic inclination oscillations with
period 196 years, coincident with TOI-178b's own oscillation period, with
maximum amplitude occurring near the 3:2 resonance location. These structures
are consistent with the system's resonant architecture and {provide a 
baseline characterization that enables future comparative studies of} similar 
phenomena in other multi-planet systems with resonant
configurations.
}


\maketitle

\section{Introduction}
The discovery and characterization of multi-planetary systems have {provided}
insights into planetary formation and evolution processes. {Among these
systems, those exhibiting mean-motion resonances (MMRs) are particularly
valuable, as they preserve information about their formation history
{\citep{Mills2016, Izidoro2017, Fabrycky2014, Lissauer2011}}. Resonant 
configurations are generally considered fragile, meaning that significant 
scattering events or giant impacts would likely disrupt the resonant chain 
{\citep{Leleu2021, Pu2015, Terquem2007}}. Thus, systems with resonant 
architectures serve as pristine laboratories for studying the outcome of 
protoplanetary discs.}

{Multi-planet systems in resonant chains represent a small but 
significant fraction of known exoplanetary systems} 
{\citep{Fabrycky2014, Wang2017}}, {with notable examples 
including TRAPPIST-1} {\citep{Gillon2017, Luger2017}}, {Kepler-90} {\citep{Gaslac2024}},
{K2-138} 
{\citep{Lopez2019}}, {Kepler-223} {\citep{Mills2016}}, 
{and HD 110067} {\citep{Luque2023}}. {The dynamics of 
small bodies in such systems have received increasing attention, with studies 
revealing complex structures analogous to those observed in our Solar System} 
{\citep{Mustill2018, Tamayo2020, Shannon2016}}.

The TOI-178 system, discovered through TESS observations and confirmed with
follow-up observations from CHEOPS, ESPRESSO, NGTS, and SPECULOOS, represents
a remarkable example of such resonant architecture \citep{Leleu2021}. This
system consists of six transiting planets orbiting a V = 11.95 mag K-dwarf
star. The planets range from super-Earth to mini-Neptune in size, with radii
between 1.1--2.9 $R_{\oplus}$ and orbital periods of 1.91, 3.24, 6.56, 9.96,
15.23, and 20.71 days {\citep{Leleu2021, Delrez2023, Leleu2024}}. 
The five outer planets form a 2:4:6:9:12 chain of Laplace resonances, while the 
innermost planet lies just outside the resonant chain \citep{Leleu2021}.

A particularly intriguing feature of TOI-178 is the non-monotonic variation
in planetary densities across the system. Unlike other known resonant systems
where density typically decreases with distance from the star, TOI-178 shows
significant variations in density between adjacent planets \citep{Leleu2021,
Leleu2024}. For instance, planet d has a significantly lower density than its
inner neighbor c, despite being closer to the star than planet e, which has
a higher density \citep{Leleu2021}. These density variations present 
challenges to standard formation and evolution models.

Recent work by \citet{Leleu2024} has refined the planetary parameters through
combined photodynamical modeling and radial velocity measurements, confirming
that the system is indeed locked in a resonant configuration with the three
Laplace angles librating around equilibrium values. The orbital structure of
TOI-178 is extremely sensitive to perturbations, with \citet{Leleu2021}
demonstrating that changing any planetary period by just $\sim$0.01 days could
result in a chaotic system.

While the orbital architecture and physical properties of the TOI-178 planets
have been extensively studied, the dynamics of test particles in this complex
resonant system remains unexplored. {Similar investigations for other
systems, such as TRAPPIST-1, have revealed rich dynamical structures including
stable and unstable regions shaped by resonances{ \citep{Quarles2020,
Unterborn2018, Grimm2018}}.} {Studies of debris dynamics in the Solar 
System have demonstrated how planetary resonances create Kirkwood gap structures.} 
{\citep{Wisdom1982, Morbidelli2002, Nesvorny2015}}, {providing 
a framework for understanding similar phenomena in exoplanetary systems}.

In this work, we investigate the dynamics of test particles in the TOI-178
system using N-body simulations. We aim to characterize
the system's dynamical architecture, identify stable and unstable regions,
{examine} the formation of resonance-induced structures analogous to Kirkwood
gaps in our Solar System, and {investigate} the stability of co-orbital regions.
{This baseline characterization provides a foundation for understanding 
small body dynamics in this multi-planet resonant chain and enables future 
comparative studies with other exoplanetary architectures.}

The remainder of this paper is organized as follows. In Section~\ref{S-numerical_simulation}, 
we describe our numerical simulation setup, including the planetary parameters, integration
methods, and initial conditions for the test particles. Section~\ref{S-results} presents the
results of our simulations, focusing on the global distribution of particles,
co-orbital region widths, resonance-induced gaps, and inclination dynamics.
Finally, in Section~\ref{S-final}, we {summarize our findings}
 and discuss the implications of our results.

\section{Numerical simulations}
\label{S-numerical_simulation}
We investigated particle dynamics in the TOI-178 system using N-body simulations with 
the \textsc{Rebound} package \citep{Rein2012}. Calculations were performed using the IAS15 
integrator \cite{Rein2015}, which employs adaptive time-stepping to maintain precision 
while preserving energy properties comparable to symplectic integrators. {We used 
the default error tolerance of IAS15 ($\sim 10^{-9}$) to ensure adequate precision for 
the close encounters and collision detection in our simulations.}

{Throughout this work, we use the terms ``test particle" and "particle" 
equivalently to refer to massless objects that serve as dynamical probes of the gravitational 
environment. These particles do not gravitationally influence the planetary 
system but respond to the combined gravitational field of the star and planets, 
allowing us to map the system's dynamical structures.}

Our simulations incorporated the six known planets of the TOI-178 system, with orbital 
elements and physical parameters adopted from \cite{Leleu2021} as summarized in 
Table~\ref{tab:planet_params}. The central star was modeled with a mass of 
$0.650$ M$_\odot$. Each planet was assigned a physical radius equal to 10\% of its 
Hill radius for collision detection purposes.
We adjusted the planetary inclinations to establish a reference plane where TOI-178c has an inclination of 0°, with all other inclinations measured relative to this plane. {Note that Table~\ref{tab:planet_params} shows the original inclinations from \cite{Leleu2021} relative to the sky plane; in our simulations, these values are adjusted such that TOI-178c serves as the reference.}

{The orbital elements were computed for the reference epoch BJD 2458741.0
following \cite{Leleu2021}.} Initial mean longitudes were
calculated from TESS observations according to

\begin{align}
\lambda_P = -\left(\frac{2\pi}{P_P}\right)(T_{0,P} - \text{date}_\text{ci}) - \frac{\pi}{2}
\end{align}

\noindent where {$\lambda_P$ is the mean longitude of planet P at the reference
epoch $\text{date}\text{ci}$, $P_P$ is the orbital period of planet P, and $T{0,P}$ is the mid-transit time
of planet P from \cite{Leleu2021}. The quantity $\text{date}\text{ci}$} represents the first observation day in BJD-TBD,
with $P_P$ and $(T{0,P} - \text{date}_\text{ci})$ expressed in consistent time units.
{The negative sign accounts for backward time propagation from the transit epoch
to the reference epoch, while the $-\pi/2$ term represents the orbital phase at which
transit occurs (when the planet crosses the observer's line of sight).}
{For all planets, the other orbital elements (argument of pericentre $\omega$,
longitude of ascending node $\Omega$, and mean anomaly $M$) were set to zero.}

\begin{table}[h]
\caption{Orbital elements and physical parameters of the TOI-178 planets.}\label{tab:planet_params}
\begin{tabular}{@{}lcccccc@{}}
\toprule
Parameter & TOI-178b & TOI-178c & TOI-178d & TOI-178e & TOI-178f & TOI-178g \\
\midrule
$a$ [au] & 0.02607 & 0.0370 & 0.0592 & 0.0783 & 0.1039 & 0.1275 \\
$e$ & 0.0035 & 0.0119 & 0.0080 & 0.0080 & 0.0105 & 0.0056 \\
$i$ [deg] & 88.8 & 88.4 & 88.58 & 88.71 & 88.723 & 88.823 \\
$M$ [M$_\oplus$] & 1.50 & 4.77 & 3.01 & 3.86 & 7.72 & 3.94 \\
$R$ [R$_\oplus$] & 1.152 & 1.669 & 2.572 & 2.207 & 2.287 & 2.87 \\
$P$ [days] & 1.91 & 3.24 & 6.56 & 9.96 & 15.23 & 20.71 \\
\botrule
\end{tabular}
\end{table}

In our primary simulation, we {distributed} $10^6$ test particles {using a 
random uniform distribution in semi-major axis (0.02 to 0.13 au) and mean anomaly 
(0 to $2\pi$). All test particles were initialized with circular orbits (eccentricity = 0) 
and the same inclination as TOI-178c (inclination = 0° in our reference frame).} 
Particles were removed from the simulation upon collision with any planet 
(approaching within the planet's physical radius) or when escaping the system (semi-major 
axis exceeding 0.3 au, approximately twice the distance of TOI-178g from the star).

We integrated the system for 500 years, with orbital parameters recorded at 10,000 equally 
spaced time points. We also conducted two additional targeted simulations 
to examine specific regions of interest. 
The first placed 10,000 test particles between 0.08 and 0.1~au (between TOI-178e and 
TOI-178f) and integrated for 500~years. The second focused on the innermost region 
(0.015-0.025 au) with 10,000 particles and extended to 10,000~years to analyze 
longer-term dynamical effects. Both simulations maintained the same planetary 
configurations and collision/escape criteria as the main simulation.

While these integration timescales are insufficient for 
full secular evolution (requiring $\sim10^8$ orbital periods), 
they are appropriate for the dynamical phenomena investigated here.
 {Co-orbital clearing occurs on
planetary clearing timescales, which can be estimated using 
the criterion presented in Eq.\ref{E-margot} below, with the most massive planet
TOI-178f requiring approximately 17,500 years to clear its neighboring
regions.}
Kirkwood gap formation develops within hundreds of years, and inclination oscillations 
complete multiple cycles within our timeframe. 
{Our 500-year integration corresponds to approximately 8,818 orbital
periods of the outermost planet TOI-178g (P = 20.71 days), providing
sufficient temporal coverage to capture the relevant clearing and resonant
phenomena.} Longer integrations with the current particle count ($10^6$)
would be computationally prohibitive for this exploratory study.


\section{Results}
\label{S-results}
Our numerical simulations of test particle dynamics in the TOI-178 system reveal
distinct structures shaped by the multi-resonant planetary configuration. We
first examine the global distribution of particles after 500 years, followed by
quantitative analysis of co-orbital regions and their deviation from theoretical
predictions. We then present evidence for resonance-induced gaps between TOI-178e
and TOI-178f, and conclude with an investigation of inclination oscillations in
the inner disk region near TOI-178b.

\subsection{Particle dynamics and resonant structures}
Our numerical simulations of test particle dynamics in the TOI-178 system reveal
distinct structural patterns shaped by the multi-resonant planetary architecture.
The gravitational influence of the six planets creates a complex landscape of
stable and unstable regions, with characteristic features associated with mean-motion
resonances. 

\begin{figure*}
    \resizebox{\hsize}{!}
    {\includegraphics{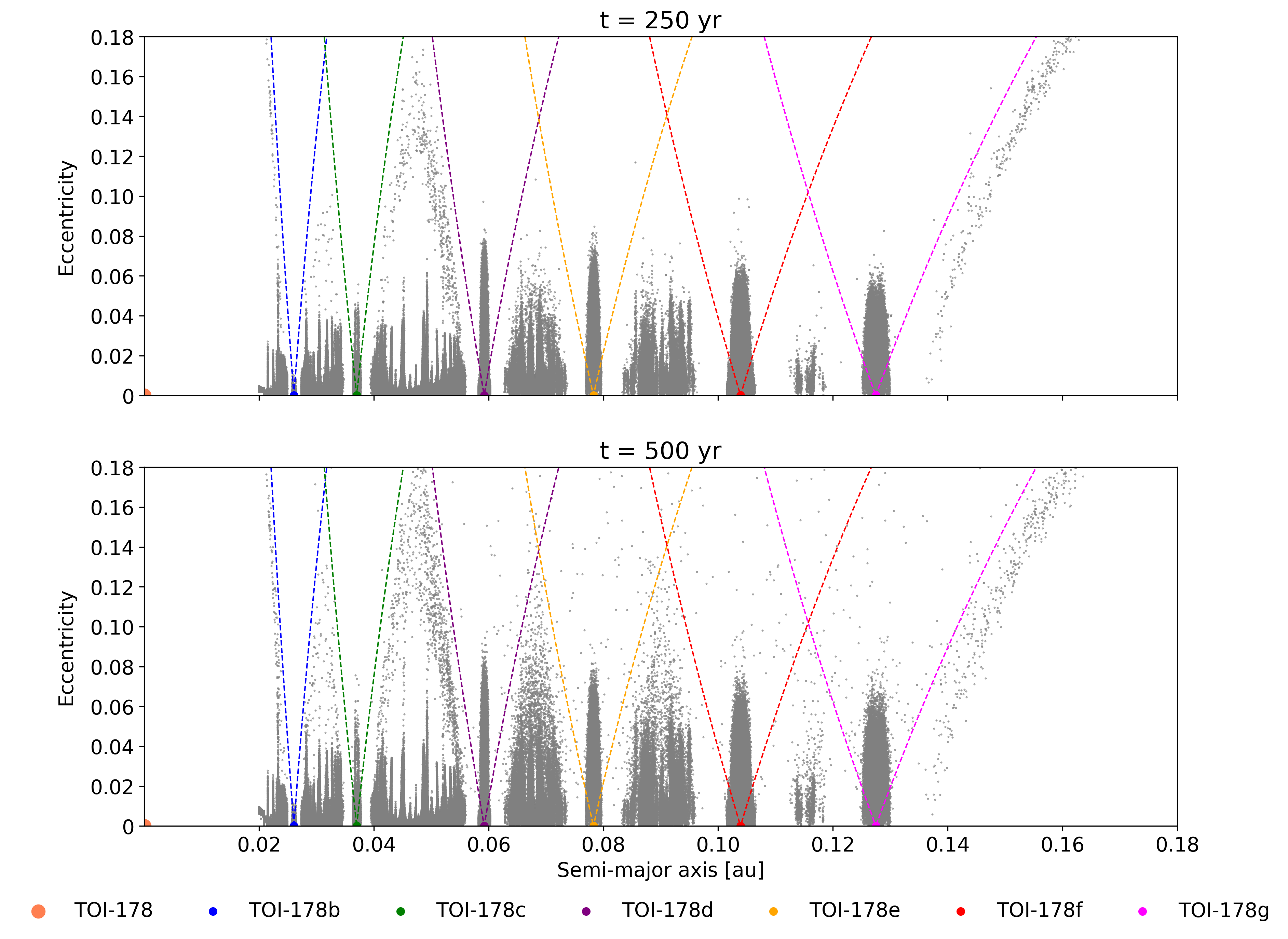}}
    \caption{Eccentricity versus semi-major axis distribution of test particles
    after 250 years (top panel) and 500 years (bottom panel), corresponding to approximately 4409 orbits and 8818 orbits for TOI-178g. Colored points mark planets, solid point at origin indicates
    TOI-178. 
    Colored dotted lines indicate orbits matching perihelion and aphelion to specific planets semi-major axes (in the same color). }
    \label{fig:ecc}
\end{figure*}

Figure \ref{fig:ecc} shows the eccentricity versus semi-major axis distribution
after 500 years. The grey points represent test particles that survived
the integration period, forming distinctive structural patterns throughout the
system. Each planet in the TOI-178 system is represented by a colored point, with
the central star shown in red at the origin. 
The dashed dotted curves indicate boundaries where
particle perihelion (left) or aphelion (right) equals the semi-major axis of each
planet.

The surviving particles predominantly occupy a confined region with semi-major axes
between 0.02~au and 0.13~au and eccentricities below 0.1. Clear vertical bands
in particle density are visible throughout this region, particularly near the
planetary orbits and at key resonance locations. These bands represent zones
where resonant perturbations have either stabilized or destabilized particle orbits,
creating a complex structure of enhanced and depleted particle densities. The
particles accumulate primarily between planetary orbits, with their distribution
bounded by perihelion and aphelion constraints that prevent close encounters with
the planets.

The dynamical evolution of particles during planetary encounters is governed by the
Tisserand relation, given by

\begin{align}
\frac{1}{2a} \sqrt{a(1-e^2)} \cos I = \frac{1}{2a'} \sqrt{a'(1-e'^2)} \cos I',
\end{align}

\noindent where primed variables denote post-encounter values. This relation, which
approximates a conserved quantity in the restricted three-body problem, explains
the characteristic boundaries in the $(a,e)$ distribution, particularly evident
beneath the aphelion line of TOI-178g at the right edge of the figure.

The efficiency with which planets clear their orbital neighborhoods depends on their
mass and orbital separation. Following \citet{Margot2016}, the clearing timescale
can be quantified as

\begin{align}
t_{\mathrm{clear}} = C^2 \, \SI{1.1e5}{yr} \left(
\frac{M_{\mathrm{star}}}{\mathrm{M}_{\odot}}\right)^{5/6} \left(
\frac{M_p}{\mathrm{M}_{\oplus}}\right)^{-4/3} \left( \frac{a}{\SI{1}{au}
}\right)^{3/2},
\label{E-margot}
\end{align}

\noindent where $C=2\sqrt{3}$ is a dimensionless constant related to the width of the cleared
zone. {The clearing timescales for the TOI-178 system range from $\sim$5800  
years for TOI-178c to $\sim30,000$ years for TOI-178g.
For context, an Earth-like planet orbiting a Sun-like star at 1~au clears a region of 
$\approx 3 R_{H,\oplus}$ of test particles in approximately $10^6$ years.}

\begin{figure*}
    \resizebox{\hsize}{!}
    {\includegraphics[width=\textwidth]{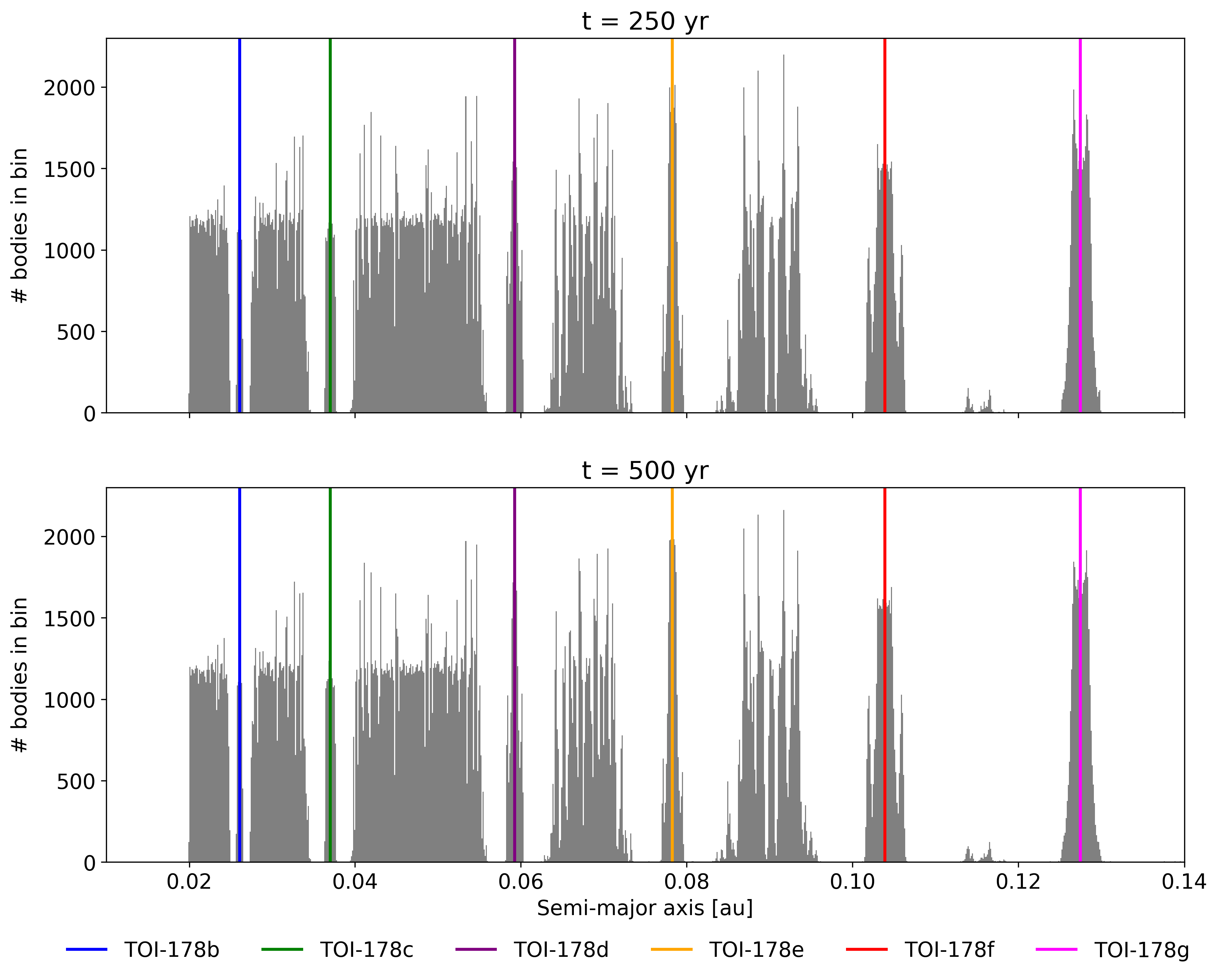}}
    \caption{Particle distribution (number of bodies in bins of size \SI{1.3e-4}{au}) evolution at 250 (top panel) and 500~years (bottom panel), corresponding to approximately 4409 orbits and 8818 orbits for TOI-178g.
    Planetary semi-major axes are indicated by solid lines.
    \label{fig-histo}}
\end{figure*}

The test particle distribution exhibits a rich resonant structure, with pronounced
vertical features at mean-motion resonances throughout the system. Figure
\ref{fig-histo} presents a histogram analysis of the particle distribution with
1000 bins from 0.02~au to 0.13~au, revealing clear gaps near planetary orbits and
co-orbital regions. Notable features include persistent high-eccentricity
populations at the 3:2 resonance of TOI-178d ($\sim$0.045~au) and the 2:1 resonance
of TOI-178f ($\sim$0.065~au). Of particular interest are the Kirkwood gap-like
structures in the region between TOI-178e and TOI-178f, visible in Figure
\ref{fig:ecc} as vertical striations at approximately 0.085~au, 0.09~au, and
0.095~au, corresponding to specific resonances with the outer planets.

\subsection{Co-orbital region widths}

The particle distribution analysis reveals distinct co-orbital zones surrounding each 
planet in the TOI-178 system. These regions, characterized by significant particle 
depletion, represent domains where the gravitational influence of each planet 
effectively clears its orbital neighborhood. To quantify the extent of these 
clearing zones, we employ a threshold-based methodology for measuring co-orbital 
widths.

Given our initial distribution of $10^6$ test particles, we establish a minimum 
threshold of 20~particles (approximately 1.5\% of typical peak values) to define 
the boundaries of each co-orbital zone. This approach allows for precise 
determination of the cleared regions despite the statistical noise inherent in 
numerical simulations. Table~\ref{tab:coorbital_comparison} presents a comprehensive comparison between 
the measured co-orbital widths and theoretical predictions based on 
\citet{Dermott1981}, whose analytical model gives:

\begin{align}
    W_{\mathrm{hs}} \sim 0.5 \mu a_{\mathrm{sat}}
\end{align}

\noindent where $W_{\mathrm{hs}}$ represents half the horseshoe width, 
$a_{\mathrm{sat}}$ denotes the satellite's semi-major axis, and $\mu$ is the mass 
ratio between the planet and the central star. This classical formula provides a 
first-order approximation for co-orbital zone widths in the circular restricted 
three-body problem.

\begin{table}[h]
    \centering
    \caption{Comparison of co-orbital region size: theory vs simulation (in [$\times 10^{-3}$\,au]).}
    \label{tab:coorbital_comparison}
    \begin{tabular}{lccc}
        \toprule
        Planet & $W_{hs,\mathrm{theo}}$ & $W_{hs,\mathrm{sim}}$ & Relative difference (\% of $W_{hs,\mathrm{sim}}$) \\
        \midrule
        TOI-178b & 0.497 & 1.04 & 52.21 \\
        TOI-178c & 1.037 & 1.56 & 33.51 \\
        TOI-178d & 1.424 & 2.21 & 35.58 \\
        TOI-178e & 2.046 & 2.86 & 28.47 \\
        TOI-178f & 3.421 & 5.07 & 32.54 \\
        TOI-178g & 3.377 & 4.94 & 31.64 \\
        \bottomrule
    \end{tabular}
\end{table}

Our results reveal significant and systematic deviations from the theoretical 
predictions. Planets within the resonance chain (TOI-178c through TOI-178g) 
exhibit co-orbital regions approximately 30\% larger than predicted by the 
\citet{Dermott1981} model, with remarkably consistent relative differences ranging 
from 28.47\% to 35.58\%. We hypothesize that this enhancement may 
be related to the complex gravitational architecture of the resonance chain. 
{Supporting this hypothesis, planets within the chain show remarkably consistent 
enhancement factors (28-36\%), while the non-chain planet TOI-178b exhibits a 
distinctly larger enhancement (52\%). This systematic difference suggests that 
different mechanisms may govern co-orbital dynamics in chain versus non-chain planets, 
though direct causal evidence requires further investigation through controlled 
simulations with varied resonant configurations.}

The most pronounced deviation occurs for TOI-178b, which shows a 52.21\% 
enhancement in its co-orbital width compared to theoretical predictions. This 
substantial increase is particularly noteworthy as TOI-178b remains outside the 
resonant chain that links the other planets. The enhanced clearing suggests that 
the standard prefactor of 0.5 in the \citet{Dermott1981} formula might 
underestimate the true co-orbital extent for non-chain planets in compact 
multi-planetary systems.

These systematic deviations highlight the limitations of simplified analytical 
models when applied to complex resonant architectures like TOI-178. 
{The consistency of enhancement factors within the resonance 
chain suggests that multi-planet resonant interactions may create more 
extensive clearing zones than predicted by two-body models, though 
this hypothesis requires systematic testing through future simulations 
with varied resonant configurations.}

\subsection{Kirkwood gap-like structures between TOI-178e and TOI-178f}

The region between TOI-178e and TOI-178f provides a detailed 
investigation of gap formation in a confirmed multi-planet Laplace resonance 
chain. Unlike the Solar System's asteroid belt, where gaps primarily result 
from Jupiter's isolated gravitational influence \citep{Kirkwood1867}, the 
TOI-178 system presents a  case where multiple planets in the 
2:4:6:9:12 resonance chain simultaneously perturb particles through 
overlapping mean-motion resonances. This configuration allows us to examine 
how resonant planetary architectures shape small body distributions. 

{Analysis of this region reveals structures analogous to the Kirkwood 
gaps but with characteristics specific to multi-planet resonant systems.} 
{The structures develop through simultaneous resonant interactions 
with both TOI-178f and TOI-178g, creating zones of depleted particle density 
at overlapping mean-motion resonances - a phenomenon not observable in 
single-planet dominated environments like our Solar System's main belt 
\citep{Morbidelli2002}.}

\begin{figure*}
    \resizebox{\hsize}{!}
     {\includegraphics[width=\textwidth]{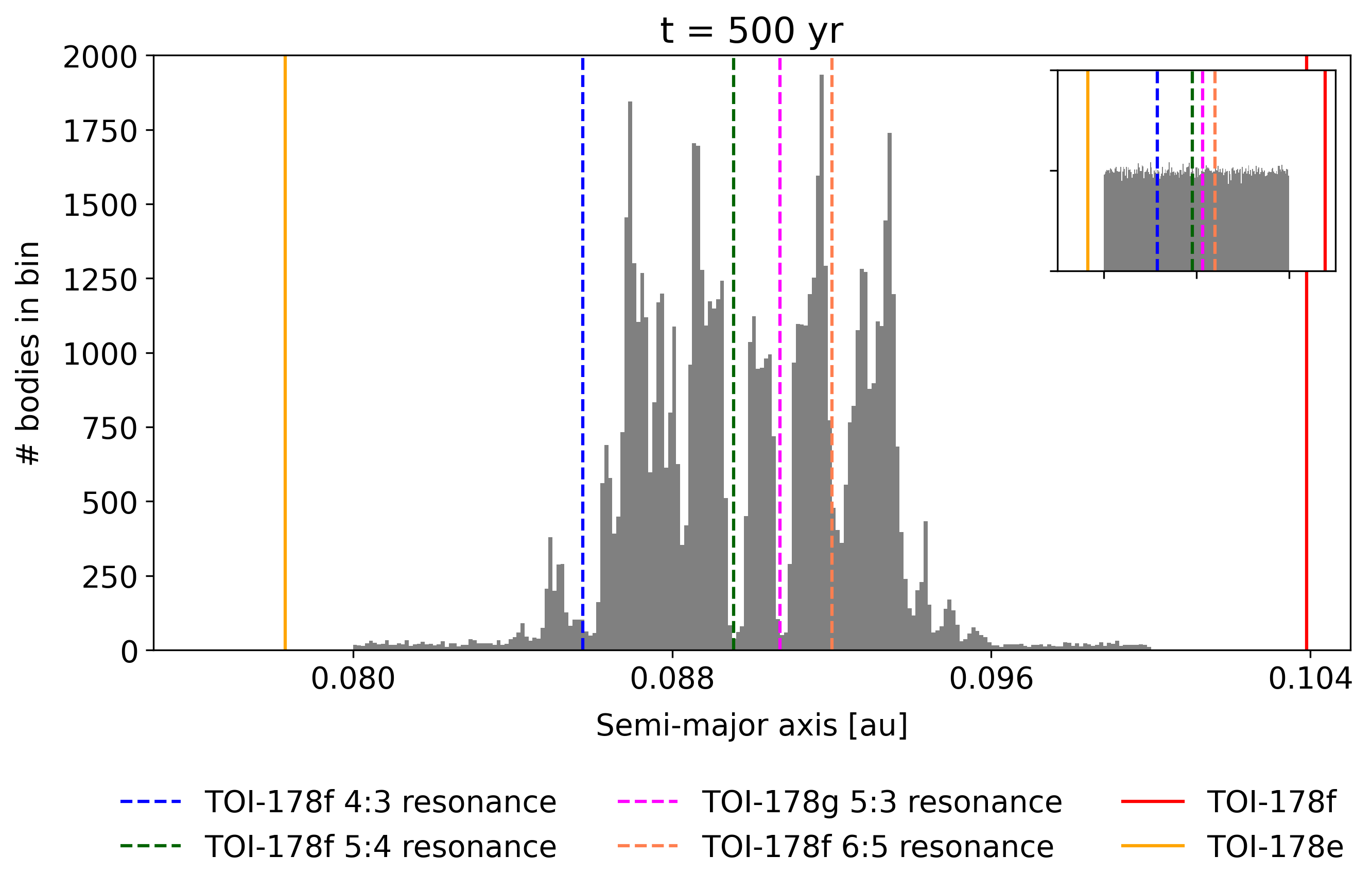}}
    \caption{Evolution of the particle distribution showing gap formation. 
    TOI-178e (yellow) and TOI-178f (red) positions and the resonances 4:3, 
    5:4, 6:5 (TOI-178f) and 5:3 (TOI-178g) are indicated. The size of one 
    bin is \SI{e-4}{au}. {The inset on the top right shows the initial distribution in the same boundaries.}}
    \label{fig-kirkwood}
    \end{figure*}

To investigate these structures in detail, we conducted a focused simulation 
with 10,000 test particles distributed between 0.08-0.1~au and tracked their 
evolution over 500~years. Figure~\ref{fig-kirkwood} shows the formation of 
prominent gaps at specific resonances: 0.089~au (5:4 resonance with 
TOI-178f), 0.091~au (5:3 resonance with TOI-178g), and 0.092~au (6:5 
resonance with TOI-178f). {Additional gaps of similar depth are visible at approximately 
0.088 au, though the specific resonant origin of this feature requires 
further investigation. We note that these gaps are not precisely centered 
at the nominal resonance locations, with typical offsets of 0.001-0.002 au 
from theoretical predictions, consistent with the finite libration widths 
and eccentricity effects in the actual system.}
A broader gap appears at 0.086~au, corresponding 
to the 4:3 resonance with TOI-178f. {The 5:3 resonance with 
TOI-178g creates gaps in a region where TOI-178f's resonances alone would not 
predict significant clearing, {suggesting the enhanced dynamical complexity}
introduced by the resonance chain architecture.} These locations align with 
mean-motion resonances where particles experience strong perturbations that 
eventually clear them from their orbits.

The resonance-driven clearing mechanism can be quantified using the libration 
width formula from resonance theory. Following \cite{Murray1999}, the maximum 
width of a resonance in semi-major axis is given by
\begin{align}
\frac{\delta a_{max}}{a} = \pm \sqrt{\left( \frac{16 |C_r|}{3 n} e\right)
\left(1+\frac{|C_r|}{27 j_2^2 e^3 n}\right)}-\frac{2 |C_r|}{9 j_2 e n},
\end{align}

\noindent where  $\delta a_{max}$ represents the maximum libration width, 
$a$ is the semi-major axis at the exact resonance location, $C_r$ denotes the 
resonance coefficient derived from the disturbing function, $n$ is the mean 
motion of the particle, $e$ is the orbital eccentricity, and $j_2$ is the 
coefficient of the mean longitude in the resonance argument.

\begin{figure*}
    \resizebox{\hsize}{!}
    {\includegraphics[width=\textwidth]{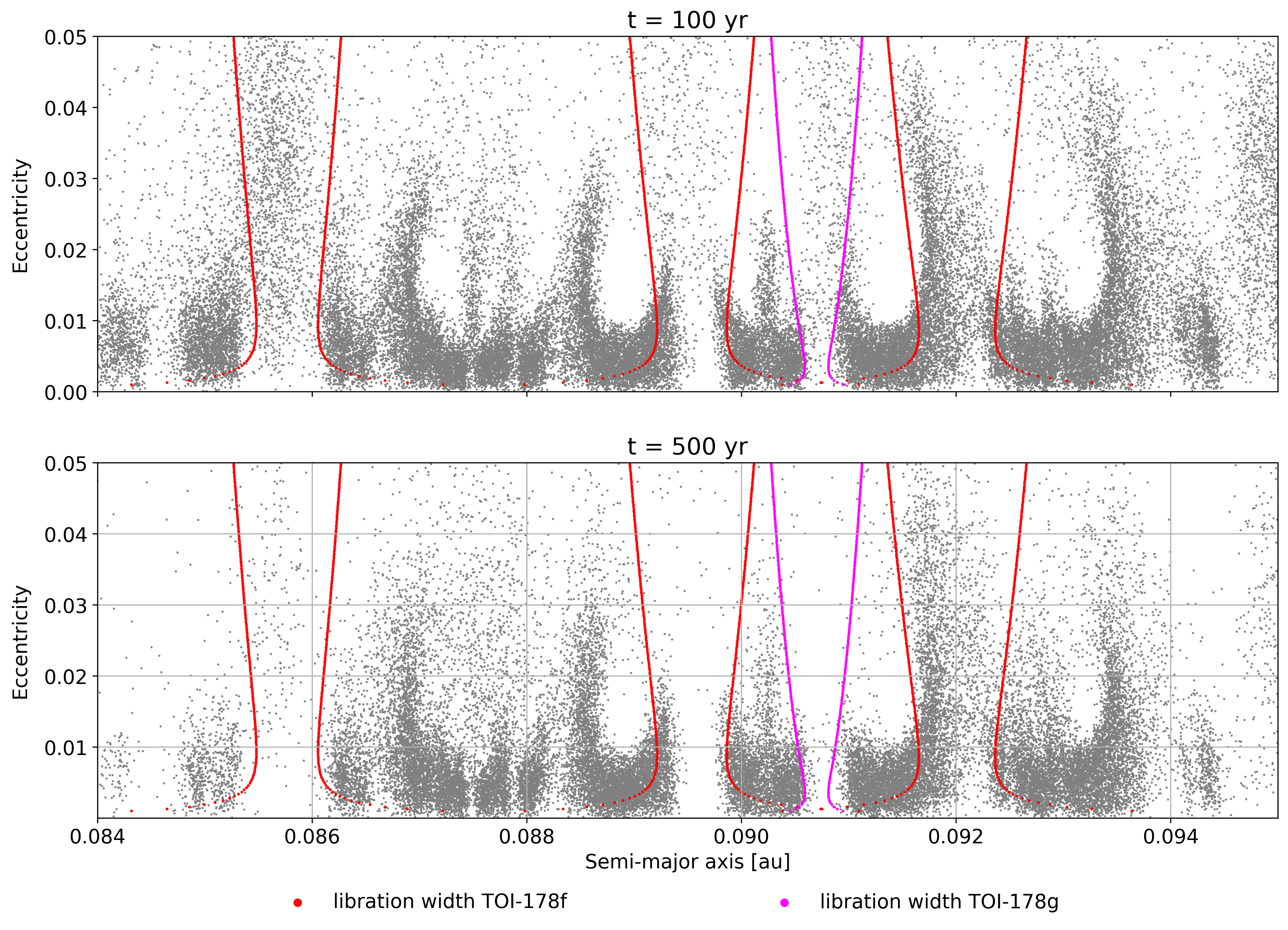}}
    \caption{Particle depletion within libration widths. Red regions: 
    TOI-178f resonances; magenta region: TOI-178g 5:3 resonance. The 
    comparison between the distribution of particles after 100~yr to 500~yr
    shows rapid zone clearing.}
    \label{fig-libration}
    \end{figure*}

Figure~\ref{fig-libration} demonstrates particle depletion within these 
theoretical libration zones. {Red regions represent resonances with 
TOI-178f, while the magenta region shows the 5:3 resonance with TOI-178g.} 
Comparing the state at 100 and 500 years reveals systematic clearing of 
resonant zones, with particles accumulating along the boundaries. The 5:4 
and 5:3 resonances exhibit particularly rapid clearing, with substantial 
depletion occurring within the first 100 years of evolution. {This 
rapid timescale (hundreds of years) may be influenced by the coherent 
gravitational perturbations from multiple planets in the resonance chain, 
as opposed to the longer clearing timescales typically associated with 
isolated planet-particle interactions \citep{Nesvorny2018}.}

{The structure formation in this system appears to be a natural 
consequence of resonant interactions between particles and the multiple 
planets, though whether the resonant chain configuration enhances or 
modifies these clearing processes compared to non-resonant multi-planet 
systems remains to be established through comparative studies.}

The width and depth of each gap correlate with the strength of the 
corresponding resonance, which depends on the mass of the perturbing planet 
and the order of the resonance. {The overlapping influence of multiple resonant planets creates gap 
patterns with enhanced complexity compared to single dominant perturbers, 
though the individual resonances remain well-described by classical theory.}
The libration width 
equation accurately predicts the locations and extents of these depleted 
regions, providing a theoretical framework for the observed clearing 
mechanisms and validating resonance theory in this multi-planet 
context. 
{While classical resonance theory accurately predicts individual gap 
locations and widths, the simultaneous presence of multiple overlapping 
resonances from different planets creates a more complex clearing pattern 
than would be expected from the simple superposition of single-planet effects.}

{These results establish baseline dynamical signatures for TOI-178's 
unique resonant architecture, filling a significant gap in the particle 
dynamics literature for this system. While extensive studies exist for 
TRAPPIST-1's resonant chain \citep{Garraffo2017,Teyssandier2022}, TOI-178's 
mathematical 2:4:6:9:12 sequence had remained unexplored from a small body 
dynamics perspective. 

\subsection{Inner disk inclination dynamics}
To investigate inclination evolution in the innermost region, we conducted an 
extended simulation with 10,000 test particles between 0.015-0.025~au, 
initialized at TOI-178c's inclination ($88.4^{\circ}$). The integration spanned 
$10^4$~y to capture long-term secular behavior, with outputs recorded at 
2000-yr intervals.

{The observed inclination oscillations arise from the natural secular 
evolution expected when test particles and TOI-178b are out of dynamical 
equilibrium} \citep{Murray1999}.

Figure \ref{fig:inc_evo_inner} shows the emergence of periodic inclination 
oscillations throughout the innermost disk region. At the beginning of the 
simulation (T=0), all particles maintain their initial inclination of 
$88.4^{\circ}$. By T=2000~yr, particles near 0.02~au begin to display 
inclination excitation, with values ranging from approximately $82^{\circ}$ to 
$94^{\circ}$. This pattern evolves further by T=4000~yr, with the amplitude of 
oscillations growing noticeably.

\begin{figure*}[h]
   \resizebox{\hsize}{!}
    {\includegraphics[width=\textwidth]{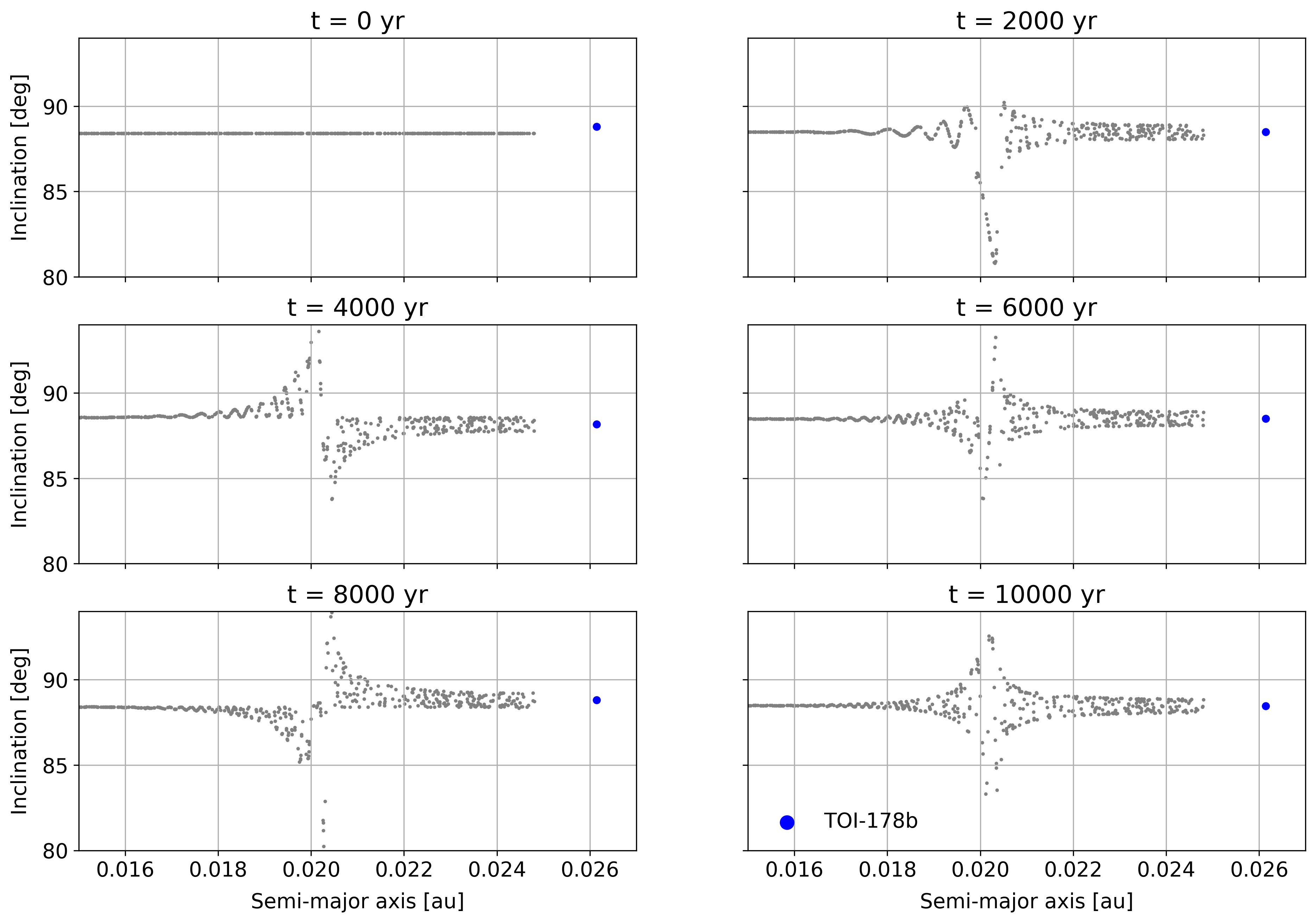}}
    \caption{Evolution of inclination showing forced oscillations. Particles are plotted in grey for six different simulation times in steps of 2000~yr and planet TOI-178b is indicated in blue.}
    \label{fig:inc_evo_inner}
\end{figure*}

The oscillation pattern exhibits a clear spatial structure. Particles with 
semi-major axes less than $\sim$0.018~au remain largely unaffected, maintaining 
inclinations close to their initial values. Between 0.018~au and 0.022~au, 
significant inclination variations develop, with the strongest oscillations 
centered at approximately 0.02~au. {This spatial concentration of 
maximum amplitude near 0.02~au is noteworthy given that the 3:2 mean motion 
resonance with TOI-178b occurs at 0.01986~au, suggesting potential resonant 
enhancement of the secular forcing mechanism.}

{The pattern cycles through different configurations with a period of 
$P_{\mathrm{inc,}b} = 196.078$~yr, which matches TOI-178b's own inclination 
oscillation period. Fourier analysis of the time series confirms TOI-178b as 
the primary forcing agent, with the power spectrum showing a dominant peak at 
the frequency corresponding to $P_{\mathrm{inc,}b}$.} {While secular 
forcing from TOI-178b naturally explains the periodic behavior, the spatial 
concentration of maximum oscillation amplitude near the 3:2 resonance location 
suggests that mean-motion resonance proximity may amplify the secular response, 
though this hypothesis requires further investigation with extended simulations.}

The T=6000~yr, T=8000~yr, and T=10000~yr panels demonstrate the continued 
stability of this oscillation pattern, with particles maintaining similar 
inclination distributions throughout the remainder of the simulation. The 
persistence of these oscillations for the full $10^4$ years, without significant 
damping, indicates that they represent a stable dynamical feature of the inner 
TOI-178 system.

This coherent inclination behavior appears unique to the innermost region, where 
TOI-178b acts as the sole significant perturber. Unlike the outer regions where 
multiple planets create complex, overlapping gravitational influences, the inner 
disk exhibits an organized, wavelike pattern characteristic of single-perturber 
systems. The sharp boundary of the oscillation zone at approximately 0.022~au 
may represent the limit of TOI-178b's dynamical influence before other planets 
begin to dominate the secular evolution.

\section{Final Remarks}
\label{S-final}

Our simulations of particle dynamics in the TOI-178 system suggest that its 
multi-resonant planetary configuration influences the structure of circumstellar 
debris. The five outer planets in the 2:4:6:9:12 resonance chain create a 
dynamical environment that shapes particle distributions in {ways that 
warrant further investigation}.

The co-orbital regions appear wider than theoretical predictions, with planets 
in the resonance chain showing about 30\% larger zones compared to 
\citet{Dermott1981} models. TOI-178b, which remains outside the chain, shows an 
even greater difference (52\%). {While the systematic difference between 
chain and non-chain planets suggests that resonant architecture may influence 
co-orbital dynamics, establishing direct causal relationships requires future 
controlled simulations with varied resonant configurations.}

Between TOI-178e and TOI-178f, we identified structures resembling Kirkwood 
gaps, providing evidence for resonance-driven clearing mechanisms. Our analysis 
shows that several mean-motion resonances, {particularly} the 5:4 with 
TOI-178f and 5:3 with TOI-178g, gradually clear particles from their libration 
zones, similar to processes seen in our Solar System's asteroid belt.

The inclination oscillations near TOI-178b reveal an interesting dynamical 
phenomenon. These oscillations, mainly found between 0.018-0.022~au with 
maximum amplitude near the 3:2 resonance location, show how a single planet can 
create organized behavior in the inner disk region. The boundary at 
approximately 0.022~au may indicate where the influence of different planets 
begins to overlap.

{This baseline characterization of} TOI-178's architecture provides a 
useful case study for resonant dynamics in exoplanetary systems. The mechanisms 
we observed -- enhanced co-orbital regions, resonance-driven gaps, and localized 
inclination patterns -- could help explain similar features in other multi-planet 
systems. 
Future comparative 
studies contrasting resonant versus non-resonant planetary configurations will 
quantify the enhancement effects suggested by this investigation.

\bmhead{Acknowledgements}
We thank the anonymous reviewers for their constructive comments and suggestions 
that significantly improved the quality of this manuscript. R.S.\ and C.M.S.\ 
acknowledge support by the DFG German Research
Foundation (project 446102036). The authors acknowledge support by the High 
Performance and Cloud Computing Group at the Zentrum für Datenverarbeitung 
of the University of Tübingen, the state of Baden-Württemberg through 
bwHPC and the DFG through grant no INST 37/935-1 FUGG. R. S. also thanks CNPq (307400/2025-5).


\section*{Declarations}
The data supporting the findings of this study, including simulation outputs and
initial conditions, are available from the corresponding author upon reasonable
request.

\bibliography{references}

\end{document}